\documentstyle[12pt]{article}
\catcode`\@=11
\long\def\@makefntext#1{
\protect\noindent \hbox to 3.2pt {\hskip-.9pt
$^{{\ninerm\@thefnmark}}$\hfil}#1\hfill}		

\def\@makefnmark{\hbox to 0pt{$^{\@thefnmark}$\hss}}  

\def\ps@myheadings{\let\@mkboth\@gobbletwo
\def\@oddhead{\hbox{}
\rightmark\hfil\ninerm\thepage}
\def\@oddfoot{}\def\@evenhead{\ninerm\thepage\hfil
\leftmark\hbox{}}\def\@evenfoot{}
\def\sectionmark##1{}\def\subsectionmark##1{}}

\renewcommand{\thefootnote}{\fnsymbol{footnote}}

\newcounter{sectionc}\newcounter{subsectionc}\newcounter{subsubsectionc}
\renewcommand{\section}[1] {\vspace*{0.6cm}\addtocounter{sectionc}{1}
\setcounter{subsectionc}{0}\setcounter{subsubsectionc}{0}\noindent
	{\normalsize\bf\thesectionc. #1}\par\vspace*{0.4cm}}
\renewcommand{\subsection}[1] {\vspace*{0.6cm}\addtocounter{subsectionc}{1}
	\setcounter{subsubsectionc}{0}\noindent
	{\normalsize\it\thesectionc.\thesubsectionc. #1}\par\vspace*{0.4cm}}
\renewcommand{\subsubsection}[1]
{\vspace*{0.6cm}\addtocounter{subsubsectionc}{1}
	\noindent {\normalsize\rm\thesectionc.\thesubsectionc.\thesubsubsectionc.
	#1}\par\vspace*{0.4cm}}

\newcounter{appendixc}
\newcounter{subappendixc}[appendixc]
\newcounter{subsubappendixc}[subappendixc]

\renewcommand{\appendix}[1] {\vspace*{0.6cm}
        \refstepcounter{appendixc}
        \setcounter{figure}{0}
        \setcounter{table}{0}
        \setcounter{equation}{0}
        \renewcommand{\thefigure}{\Alph{appendixc}.\arabic{figure}}
        \renewcommand{\thetable}{\Alph{appendixc}.\arabic{table}}
        \renewcommand{\theappendixc}{\Alph{appendixc}}
        \renewcommand{\theequation}{\Alph{appendixc}.\arabic{equation}}
        \noindent{\bf Appendix \theappendixc #1}\par\vspace*{0.4cm}}

\def\abstracts#1{{

\centering{\begin{minipage}{12.2truecm}\footnotesize\baselineskip=12pt\noindent
	\centerline{\footnotesize ABSTRACT}\vspace*{0.3cm}
	\parindent=0pt #1
	\end{minipage}}\par}}

\renewenvironment{thebibliography}[1]
	{\begin{list}{\arabic{enumi}.}
	{\usecounter{enumi}\setlength{\parsep}{0pt}
\setlength{\leftmargin 1.25cm}{\rightmargin 0pt}
	 \setlength{\itemsep}{0pt} \settowidth
	{\labelwidth}{#1.}\sloppy}}{\end{list}}

\newcounter{itemlistc}
\newcounter{romanlistc}
\newcounter{alphlistc}
\newcounter{arabiclistc}

\newcommand{\fcaption}[1]{
        \refstepcounter{figure}
        \setbox\@tempboxa = \hbox{\footnotesize Fig.~\thefigure. #1}
        \ifdim \wd\@tempboxa > 6in
           {\begin{center}
        \parbox{6in}{\footnotesize\baselineskip=12pt Fig.~\thefigure. #1}
            \end{center}}
        \else
             {\begin{center}
             {\footnotesize Fig.~\thefigure. #1}
              \end{center}}
        \fi}

\newcommand{\tcaption}[1]{
        \refstepcounter{table}
        \setbox\@tempboxa = \hbox{\footnotesize Table~\thetable. #1}
        \ifdim \wd\@tempboxa > 6in
           {\begin{center}
        \parbox{6in}{\footnotesize\baselineskip=12pt Table~\thetable. #1}
            \end{center}}
        \else
             {\begin{center}
             {\footnotesize Table~\thetable. #1}
              \end{center}}
        \fi}

\def\@citex[#1]#2{\if@filesw\immediate\write\@auxout
	{\string\citation{#2}}\fi
\def\@citea{}\@cite{\@for\@citeb:=#2\do
	{\@citea\def\@citea{,}\@ifundefined
	{b@\@citeb}{{\bf ?}\@warning
	{Citation `\@citeb' on page \thepage \space undefined}}
	{\csname b@\@citeb\endcsname}}}{#1}}

\newif\if@cghi
\def\cite{\@cghitrue\@ifnextchar [{\@tempswatrue
	\@citex}{\@tempswafalse\@citex[]}}
\def\citelow{\@cghifalse\@ifnextchar [{\@tempswatrue
	\@citex}{\@tempswafalse\@citex[]}}
\def\@cite#1#2{{$\null^{#1}$\if@tempswa\typeout
	{IJCGA warning: optional citation argument
	ignored: `#2'} \fi}}

 1
 1
 1

\font\ninerm=cmr9

\begin{document}

\hfill
{\vbox{
\hbox{CPP-95-4}
\hbox{DOE-ER-40757-064}
\hbox{March 1995}}}

\centerline{\normalsize\bf AN UPDATE ON STRONG $W_L W_L$ SCATTERING AT THE LHC
\footnote{Talk presented at Beyond the Standard Model IV,
Lake Tahoe, California (Dec 1994)}
}
\centerline{\footnotesize KINGMAN CHEUNG\footnote{Representing also J. Bagger,
V. Barger, J. Gunion, T. Han, G. Ladinsky, R. Rosenfeld, and C.P. Yuan}}
\baselineskip=13pt
\centerline{\footnotesize\it
Center for Particle Physics, University of Texas at Austin,
Austin TX 78712, U.S.A.}
\centerline{\footnotesize E-mail: cheung@utpapa.ph.utexas.edu}

\vspace*{0.9cm}
\abstracts{
I summarize an update on the study for a strongly interacting
electroweak symmetry breaking sector via longitudinal vector boson scattering
at the 14 TeV Large Hadron Collider.
In the update,  the decay mode $ZZ\to \ell^+\ell^- \nu \bar \nu$ and
a new vector-resonance signal via $q\bar q' \to V
\to W^+W^-/ W^\pm Z$  are also included.
}

\normalsize\baselineskip=15pt
\setcounter{footnote}{0}
\renewcommand{\thefootnote}{\alph{footnote}}

\vspace{0.2in}

In a recent paper \cite{1} we presented a thorough signal-background
analysis of the strongly-interacting electroweak symmetry breaking sector
(SEWS) via longitudinal vector boson scattering, in which the gold-plated
decay modes of $W$ and $Z$ bosons are considered.
But that paper emphasized on the senate-killed SSC parameters and adopted
a similar set of acceptance cuts for the 16 TeV LHC.  Now we know that
the LHC has been approved with an energy of 14 TeV.   Since the
signal of the SEWS is rather sensitive to the energy of the machines,
we have performed an updated analysis \cite{2} to optimize the acceptance
cuts for the LHC with the updated energies and luminosities
(100 fb$^{-1}$ per year).  I summarize the update here.

Our signal of interest mainly comes from $W_L W_L$ fusion:
\begin{equation}
q q' \to qq' Z_L Z_L, qq'W^+_L W^-_L, qq'W^\pm_L Z_L, qq'W^\pm_L W^\pm_L \;,
\end{equation}
followed by $W\to \ell \nu$ and $Z\to \ell^+\ell^-$ decays.
The strategies to extract the $W_L W_L$ signals from the Standard Model
(SM) backgrounds follow closely as in Ref.~1.  In the update,
we also extended to include the $ZZ\to \ell^+\ell^- \nu \bar \nu$ decay
mode to supplement the $ZZ\to \ell^+\ell^-\ell^+\ell^-$ channel, which
has less than 10 events per year.  We have also included the
$q \bar q' \to W^* \to V \to W^+_L W^-_L/W_L Z_L$ processes for a
vector resonance $V$ via $W-V$ mixing, which has been
proved more useful in searching for the vector resonance than
the $W_L W_L$ fusion mechanism.  For background processes we also
used the value of $m_t=175$~GeV and all the $q\bar q'\to WW$ + QCD jets
are reevaluated to include ${\cal O}(\alpha_s)$ corrections.
In addition, we have also included one detector-dependent background,
$W^+Z\to \ell^+ \ell^+$ to the $W^+_L W^+_L$ channel when the $\ell^-$
{}from the $W^+Z$ decay escapes outside the detector range.

As described in our earlier paper \cite{1}
we again consider seven models for the SEWS physics:
(i) the ``SM'' with $m_H=1$ TeV;
(ii) the ``Scalar'' model with a spin-0, isospin-0 chirally coupled
resonance with mass of 1 TeV and width 350 GeV;
(iii) the ``$O(2N)$'' model with $N=2$ and
an amplitude having a pole at $s=[m-i\Gamma/2]^2$ with
$m=0.8$ TeV and $\Gamma=600$ GeV;
(iv) a ``Vector'' model with a spin-1, isospin-1
chirally coupled resonance;  we choose the mass-width combinations
as ($M_V,\Gamma_V$)=(1 TeV,5.7 GeV) and (2.5 TeV,520 GeV);
(v) the non-resonant ``LET-CG'' model of Chanowitz and Gaillard
in which the low energy theorem (LET) amplitude is used
and the unitarity saturation is assumed once  the partial waves
reach the unitarity bound;
(vi) the non-resonant  ``LET-K'' model in which
the LET  amplitude is used
and the unitarization of the partical waves is achieved by K-matrix;
(vii) the ``Delay-K'' model in which one-loop correction terms
to the LET amplitude are chosen so as to delay the onset
of unitarity violation to energies beyond 2 TeV, and K-matrix unitarization
is employed to ensure unitarity beyond this point.

In the following I shall describe the different characteristics between
the signal and various backgrounds, which include
the electroweak production of transverse $WW$ pair, the lowest order
production of $WW$ pair in association with QCD jets, and top-related
backgrounds.  Since the scale of the SEWS is of order TeV, the $W_L W_L$
scattering via the dynamics of the SEWS is characterized by several
unique features that are quite different from the backgrounds:
\begin{description}
\item[(i)] the leptons coming from
the decays of the $W_L$ and $Z_L$ after strong
scattering are very energetic and very back-to-back in the transverse
plane.  These features prompt us to consider high $p_T$, central
rapidity and large invariant mass cuts, as well as large
$\Delta p_T(\ell\ell)$ and $\cos \phi_{\ell\ell}$ cuts;

\item[(ii)] the presence of very energetic ($\sim 1$ TeV), small $p_T$, and
forward jets in association with the $W_L W_L$ fusion.  This motivates us
to tag forward energetic jets, which is especially effective in reducing
the $WW+$ QCD jet backgrounds;

\item[(iii)] the absence of large $p_T$ jets in the central rapidity region.
This prompts us to veto any hard jets in the central region.  This is
extremely effective in suppressing the top-related backgrounds and the
EW backgrounds.
\end{description}

Specifically, we started with the model of 1 TeV SM Higgs boson because it
can be incorporated consistently into the SM simply by setting $m_H=1$ TeV,
namely, the signal of the 1 TeV Higgs boson is defined as
$\sigma(SM\; m_H=1\;{\rm TeV})-\sigma(SM\; m_H=0.1\;{\rm TeV})$,
where $\sigma(SM\; m_H=0.1\;{\rm TeV})$ represents the EW background.
We then came up with a set of optimized cuts and the jet-veto and
jet-tag efficiencies.
The specific cuts and jet efficiencies used in different channels
can be found in Ref.~2.
Since not all these SEWS models can be incorporated into the SM consistently,
we employed the Effective-$W$ Approximation (EWA)
in combination with the Equivalence Theorem (ET) to calculate the signal rates.
In this EWA/ET approach we first compute  cross sections ignoring
all jet observables but implementing all the leptonic cuts.
Then to obtain the cross sections that include
the jet-tagging and jet-vetoing cuts, we  simply multiply
by the jet-tag and/or jet-veto efficiencies
as obtained for the SM 1 TeV Higgs boson signal.
This procedure is justified since the kinematics of the jets in the
signal events are determined only by the initial $W_L$'s
and therefore should be independent of the SEWS dynamics.
\begin{table}[t]
\caption[]{\small Event rates per LHC year for (a) the $W_L W_L$ fusion signals
for various SEWS models in channels of vector boson pair,
and (b) for $q\bar q\to W^+ W^-$ and $q \bar q' \to W^\pm Z$ channels
deriving from $W-V$ mixing. \label{table1}
}
\bigskip
\centering
\begin{tabular}{l@{\extracolsep{-0.064in}}|ccccccccc}
\hline
\hline
(a) & Bkgd. & SM & Scalar & $O(2N)$ & Vec 1.0 & Vec 2.5 & LET-CG &
   LET-K & Delay-K \\
\hline
$Z Z(4\ell)$ & 0.7 & 9 & 4.6 & 4.0 & 1.4 & 1.3 & 1.5 & 1.4 & 1.1  \\
 \hline
$Z Z(2\ell2\nu)$ & 1.8 & 29 & 17 & 14 & 4.7 & 4.4 & 5.0 & 4.5 & 3.6  \\
 \hline
$W^+W^-$ & 12 & 27 & 18 & 13 & 6.2 & 5.5 & 5.8 & 4.6 & 3.9 \\
  \hline
$W^\pm Z$ & 4.9 & 1.2 & 1.5 & 1.2 & 4.5 & 3.3 & 3.2 & 3.0 & 2.9 \\
  \hline
$W^\pm W^\pm$ & 3.7 & 5.6 & 7.0 & 5.8 & 12 & 11 & 13 & 13 & 8.4 \\
\hline
\end{tabular}

\bigskip

\begin{tabular}{l@{\extracolsep{0.1in}}|ccc}
\hline
\hline
(b) & Bkgd. & Vec1.0: $W$-$V$mix / fusion & Vec2.5: $W$-$V$ mix /  fusion \\
\hline
$W^+ W^-$ & 420 & 8.6 / 10 & 0.3 / 9.0  \\
\hline
$W^\pm Z$ & 220 & 73 / 8.7 & 1.4 / 6.4  \\
\hline
$W^\pm Z$ & & $0.85<M_T<1.05$ TeV      &  $2<M_T<2.8$ TeV  \\
B/mix/fusion &   & 22/ 69  / 3.2 &   0.82/0.81/0.55 \\
\hline
\end{tabular}
\end{table}
The final numbers for the $W_LW_L$ fusion
signals and backgrounds for various channels
are summarized in Table~\ref{table1}(a).
Large excesses above SM backgrounds
are predicted in $ZZ(4\ell)$, $ZZ(2\ell2\nu)$, and $W^+ W^-$
modes for scalar-type models; the vector-type models would yield
observable event excess in the $W^\pm W^\pm$ channel,
but to a much less extent in the $W^\pm Z$ channel; whereas
the non-resonant models yield observable excesses in the $W^\pm W^\pm$ channel.
Therefore, an observation of excess vector boson pairs in a particular
channel will signal a specific dynamics of SEWS.
On the other hand, the
 vector resonance can also be probed via the Drell-Yan process
$q\bar q' \to W^* \to V \to W^\pm Z, W^+ W^-$, which  are more
important as long as $V$ is not too heavy.
However, we have to drop the jet-tag cut because the Drell-yan processes
do not have accompanying jets at the lowest order.
We calculate the signal assuming 100\%  jet-veto efficiency.
The resulting signal and background event rates are shown in
Table~\ref{table1}(b).
Despite the increase in backgrounds due to dropping the jet-tag,
the increase in signal event rate for a 1 TeV vector
resonance presents a clear bump near the resonance mass in
the $M_T$ spectrum.
The $q \bar q' \to W^\pm Z$ channel  via  $W$-$V$ mixing should be the
best to study
a Vector resonance model at the LHC if $M_V \sim 1$ TeV, but
for a 2.5 TeV vector state the signal rates are too small for any practical
detection; whereas the $q\bar q \to W^+W^-$ is less useful in probing
the vector resonance due to enormous backgrounds.
This work was supported by DOE-FG03-93ER40757.


\begin{thebibliography}{99}
\bibitem{1}J. Bagger {\it et al.}, Phys. Rev. {\bf D49} (1994) 1246.
%
\bibitem{2}J. Bagger {\it et al.}, preprint CPP-95-3
( March 1995).
\end{thebibliography}
\end{document}